\documentclass[aps,prx,onecolumn,amsmath,nofootinbib,amssymb,11pt]{revtex4-1}
\usepackage{graphicx}
\usepackage{txfonts}
\usepackage[colorlinks, linkcolor=blue]{hyperref}
\usepackage{verbatim}
\usepackage{esint}
\usepackage{amsmath}
\usepackage{hyperref}
\usepackage{mathtools}
\renewcommand{\vec}[1]{\boldsymbol{#1}}

\def \a{{\vec a}}

\def \e {{\vec e}}
\def \j {\vec{j}}
\def \E {{\vec E}}
\def \vp {\varphi}
\def \r {{\vec r}}

\def \s {\psi}
\def \q {{\vec q}}
\def \d{\partial}

\def \ve {\varepsilon}

\def \z {\zeta}

\def \D{\Delta}

\def \A{{\bf A}}

\def \L{{\cal{L}}}

\def \beq {\begin{eqnarray}}
\def \eeq {\end{eqnarray}}
\def \tn {\textnormal}

\def \M {\vec {M}}

\def \M {M\overline{M}}

\def \ua{\uparrow}
\def \da{\downarrow}

\def \scf{\sigma_\text{CF}}
\def \rhocs{\hat{\rho}_\text{CS}}
\def \ecs{\mathbf{e}}
\def \kf {k_\text{F}}
\def \lb {\ell_B}

\begin{document}
\title{Semiclassical theory of the tunneling anomaly in partially spin-polarized compressible quantum Hall states}
\author{Debanjan Chowdhury}
\thanks{These two authors contributed equally.}
\author{Brian Skinner}
\thanks{These two authors contributed equally.}
\author{Patrick A. Lee}
\affiliation{Department of Physics, Massachusetts Institute of Technology, Cambridge Massachusetts
02139, USA.}
\date{\today \\
\vspace{.1in}}

\begin{abstract}

Electron tunneling into a system with strong interactions is known to exhibit an anomaly, in which the tunneling conductance vanishes continuously at low energy due to many-body interactions.  Recent measurements have probed this anomaly in a quantum Hall bilayer of the half-filled Landau level, and shown that the anomaly apparently gets stronger as the half-filled Landau level is increasingly spin polarized. Motivated by this result, we construct a semiclassical hydrodynamic theory of the tunneling anomaly in terms of the charge-spreading action associated with tunneling between two copies of the Halperin-Lee-Read state with partial spin polarization. This theory is complementary to our recent work (\href{https://arxiv.org/abs/1709.06091}{{arXiv:1709.06091}}) where the electron spectral function was computed directly using an instanton-based approach. Our results show that the experimental observation cannot be understood within conventional theories of the tunneling anomaly, in which the spreading of the injected charge is driven by the mean-field Coulomb energy.  However, we identify a qualitatively new regime, in which the mean-field Coulomb energy is effectively quenched and the tunneling anomaly is dominated by the finite compressibility of the composite Fermion liquid.
\end{abstract}
\maketitle

\tableofcontents

\section{Introduction}

In an interacting two-dimensional electron system, the amplitude for tunneling an additional electron into the system is influenced not just by the single-particle density of states but also by the strength of electron-electron interactions. This influence is particularly strong when the energy of the injected electron (relative to the Fermi level) is low compared to the typical scale of electron-electron interactions.  At such low energies, inserting an electron requires other nearby electrons to rearrange, clearing out a ``correlation hole" into which the tunneled electron can be placed. For systems with sufficiently strong interactions and with finite conductivity, this many-body interaction effect leads to a ``tunneling anomaly" (TA), in which the tunneling conductance vanishes as the bias voltage is brought to zero. \cite{PAL80}

\begin{figure}
\begin{center}
\includegraphics[width=0.7\columnwidth]{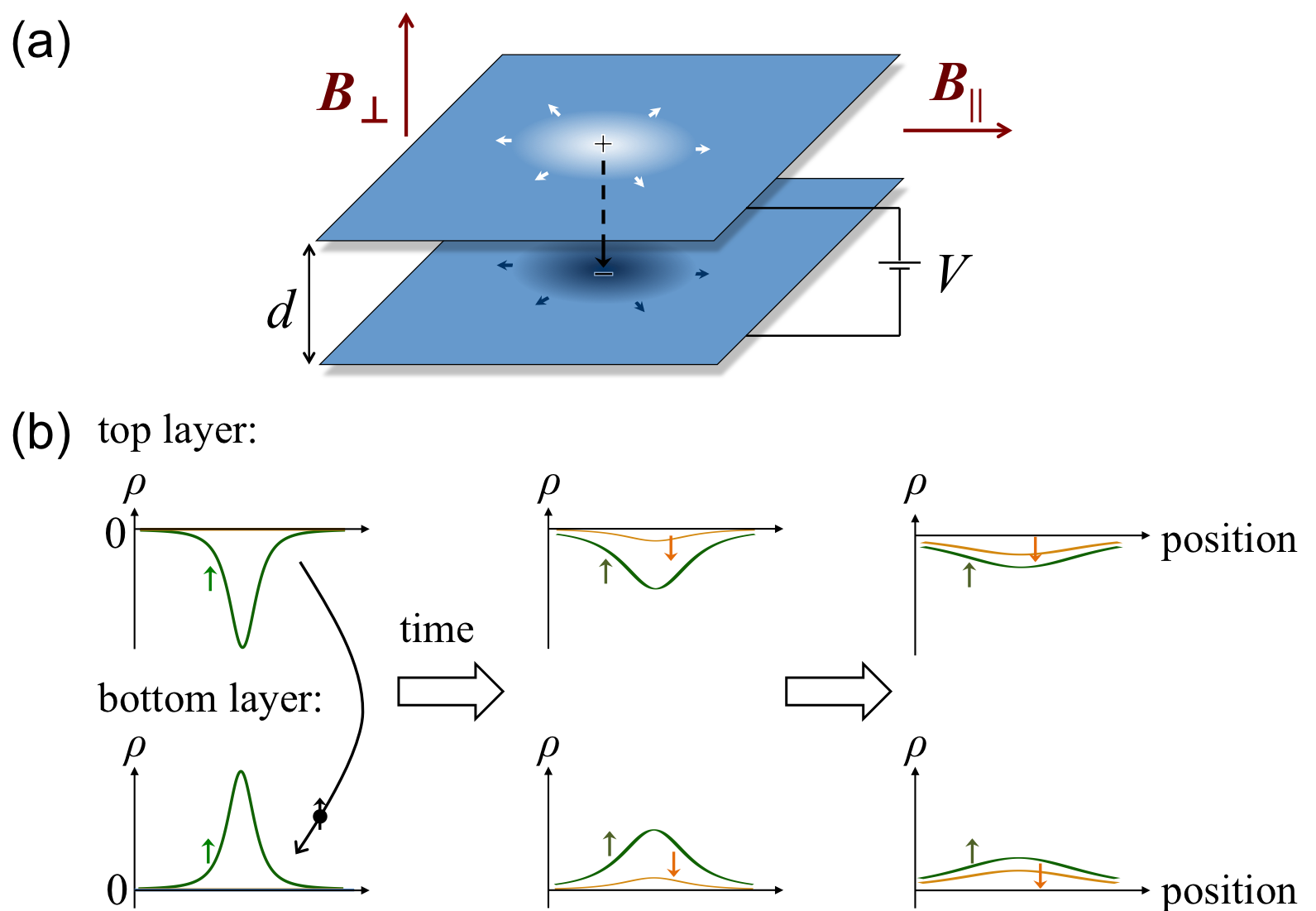}
\end{center}
\caption{{\bf (a)} Schematic depiction of the charge-spreading process. Two parallel two-dimensional electron systems are separated by a distance $d$, and subjected to in-plane and out-of-plane magnetic fields, $B_\parallel$ and $B_\perp$, respectively.  A bias voltage $V$ drives tunneling of electrons from the top layer to the bottom layer. Such tunneling processes locally perturb the CF density (indicated by the blue color of varying intensity). One can describe the rate of electron tunneling by considering the action associated with the outward spreading of the positive charge left behind in the top layer (light-colored area, indicating a lower density of CFs) and the negative charge deposited in the bottom layer (dark-colored area, indicating a higher density of CFs). {\bf (b)} Evolution of the spin composition of the spreading charge. Immediately after tunneling of a spin up electron, the top layer has a local deficit of up-spin CF density, while the bottom layer has an excess. As the charge density perturbation spreads outward, the Chern-Simons electric field mixes the two spin components within each layer, so that after a long time the ratio of spin-up to spin-down density in the spreading charge perturbation, $ \rho_\uparrow / \rho_\downarrow$, is determined solely by the thermodynamic compressibilities of each spin species, and is independent of the spin of the tunneled electron.}
\label{cartoon}
\end{figure}

Conceptually, one can think that the tunneling process comprises two distinct steps: (i) a fast, single-particle transmission of an electron across the tunneling barrier, and (ii) a slow, many-body rearrangement of the electron liquid in response to the transmitted electron. At low bias voltage, the latter process acts as a bottleneck that determines the tunneling rate.  One can describe step (ii) using the language of ``charge spreading".  In this picture, the additional charge density associated with the injected electron is effectively spread outward by the rearrangement [as depicted in Fig.\ \ref{cartoon}(a)], reducing the system energy closer to that of the ground state.  At zero temperature, the charge spreading can happen only as a virtual process, and therefore the tunneling rate is proportional to $\exp(-S/\hbar)$, where $S$ is the action associated with charge spreading.  In general, the action $S$ grows with decreasing bias voltage $V$, since the charge of the tunneled electron must spread far enough outward during the virtual process that the change in the system energy is reduced below $eV$.  (Here, $-e$ is the electron charge.)  A number of system properties are reflected in the magnitude of the charge spreading action, including the interaction strength, the conductivity, and the electronic compressibility, and therefore the tunneling anomaly can generally be used as a probe of the many-body ground state.

Of particular interest in the study of correlated electron systems are the compressible states that exist at even-denominator quantum Hall filling fractions.  Such states have been successfully described using the framework of composite fermions (CFs), in which each electron is attached to an even number $\phi$ of flux quanta \cite{JKJ}.  For example, the state at filling factor $\nu = 1/2$ was described by Halperin, Lee, and Read (HLR) \cite{HLR} in terms of a low-energy effective theory for the CFs coupled to an emergent Chern-Simons (CS) gauge field.  In a compressible quantum Hall state, the filling is such that all of the system's magnetic field is bound up in the CFs, so that the CFs effectively see zero magnetic field on average and form a Fermi surface. When the tunneling anomaly is measured in such a compressible quantum Hall state, it is this Fermi liquid of CFs whose properties are probed.  That is, the charge-spreading is accomplished by rearrangement of the CF Fermi sea, and the charge-spreading action $S$ reflects the conductivity, the interaction strength and the finite compressibility of the CFs.  

Tunneling into quantum Hall systems has attracted experimental and theoretical interest for almost thirty years.  For example, experiments have shown clear evidence for a TA in quantum Hall systems \cite{Ashoori, Eis92, Jones94}. The problem of the tunneling anomaly in the $\nu = 1/2$ state has received particular theoretical attention, \cite{HPH, XGW94} with previous authors assuming complete spin polarization and focusing primarily on the case of a single layer.  A recent experiment \cite{JPE16}, however, has re-examined the problem of tunneling anomaly in the $\nu = 1/2$ state by measuring the tunneling current between two closely-spaced GaAs quantum wells that are each at half filling.  Crucially, this experiment also examined the role of partial spin polarization by varying an in-plane magnetic field $B_\parallel$.  Numerous studies during the past two decades have shown that at low electron density the $\nu = 1/2$ state in GaAs is not fully polarized \cite{Kukushkin, Dementyev, Shayegan, Spielman, Kumada, Dujovne, Tracy,  Giudici, Smet1, Smet2, Finck}.  Applying an additional in-plane field allows one to increase the spin polarization without affecting the orbital state of the electrons.  

One of the most striking observations of Ref.\ \onlinecite{JPE16} is that the tunneling current at low bias \emph{decreases} with increasing spin polarization.  As we show below, this observation is at odds with conventional treatments of the tunneling anomaly, which predict a tunneling current that increases with spin polarization.  This experimental discrepancy prompts us to revisit the theory of the tunneling anomaly in bilayers of compressible quantum Hall states.  In doing so we uncover a new regime of behavior for the tunneling anomaly, in which the charge-spreading action is dominated not by the mean-field Coulomb energy of the spreading charge but by the finite compressibility of the CF liquid.  In this regime the dependence of the tunneling anomaly on spin polarization depends on the behavior of the compressibility, which is strongly renormalized by interactions.  

In our description we focus everywhere on the limit of low voltages $V \ll e/(\epsilon \lb)$, where $\epsilon$ is the dielectric constant and $\lb$ is the magnetic length. (We use Gaussian units throughout this paper.) At such low voltages the current is far below its maximum value $I_\text{max}$ and the physics associated with charge spreading over distances $\gg \lb$ plays a dominant role.  The behavior of the current peak was explored in Ref.\ \onlinecite{JKJ17}.  The authors found that the evolution of the peak with in-plane field can be explained in terms of the momentum boost given to the tunneling electron, which produces a lateral shift in the position of the electron's guiding center.  At low voltages this shift is much smaller in magnitude than the typical radius of the spreading charge, and is not relevant for the problem we are considering.  We also neglect everywhere the possibility that electrons and holes in opposite layers couple to form an exciton condensate (reviewed, for example, in Refs.\ \onlinecite{EisReview, excrev}); in the regime of our interest there is no experimental evidence for an excitonic condensate, which produces a zero-bias peak in the tunneling conductance. Our assumption of no excitonic coupling between layers is equivalent to assuming that either the ratio $d/\lb$ is larger than the critical value associated with exciton condensation, or that the temperature is larger than the condensation temperature.

In the remainder of this paper we calculate the zero-temperature charge-spreading action as a function of bias voltage $V$ and inter-layer separation $d$.  In Sec.\ \ref{sec:qualitative} we first define our model and then provide a semi-quantitative derivation of our main results.   In Sec.\ \ref{sec:hydro} we present a calculation of the charge-spreading action using a semiclassical hydrodynamic description, following Ref.\ \onlinecite{LS}. We discuss its implications for the tunneling anomaly and its polarization dependence across different regimes of $V$ and $d$. In Sec.\ \ref{sec:instanton} we briefly review the calculation of the single-electron spectral function, as was done recently by us in an accompanying paper \cite{DCBSPL1}, and we compare our results to the hydrodynamic approach. The two descriptions give equivalent results.  We conclude in Sec.\ \ref{sec:discussion} with suggestions for future experiments and a brief theoretical outlook.

\section{Preliminaries and semi-quantitative discussion}
\label{sec:qualitative}

In this paper we consider the situation where two parallel layers have the same overall electron concentration $n$ and polarization $\zeta$, defined such that
\beq
n_\ua + n_\da & = n,		\nonumber \\
n_\ua - n_\da & = \zeta n , 
\label{glob}
\eeq
where $n_\ua$ and $n_\da$ denote the overall concentration of up and down spin electrons, respectively, in each layer. We are interested in the situation where we attach an even number, $\phi$, of flux quanta to the electrons such that each resulting composite Fermion sees an average effective magnetic field of zero.  This is accomplished by attaching flux in such a way that each CF sees $\phi$ flux quanta attached to electrons of either spin component in the same layer, and no flux quanta attached to electrons in the opposite layer \cite{NB93}.  Since the average effective magnetic field for CFs is $B_\perp - 2 \pi \phi n/e = 0$, our description applies to filling factors $\nu = 2 \pi n \lb^2 = 1/\phi$. Here $\lb = \sqrt{\hbar c/e B_\perp}$ denotes the magnetic length and $B_\perp$ is the magnetic field perpendicular to the layers.

Within the language of HLR, at incomplete spin polarization each layer houses two Fermi surfaces with different Fermi wave vectors 
\beq
k_{\text{F}\ua} &=& \sqrt{4 \pi n_{\ua}} = \kf \sqrt{\frac{1+\z}{2}} \nonumber \\
k_{\text{F}\da} &=& \sqrt{4 \pi n_{\da}} = \kf \sqrt{\frac{1-\z}{2}},
\label{kf}
\eeq
where $\kf = \sqrt{2 \nu}/\lb$ is the Fermi wave vector in the limit of complete spin polarization ($\zeta=1$).  

In describing the charge spreading, we use the symbol $\rho_{\sigma, s}(\r, t)$ to denote the spatially- and temporally-varying differential charge density, defined relative to the uniform background $en_\sigma$, associated with CFs having spin $\sigma ~(=\ua, \da)$ in layer $s~ (=1,2)$.
When an electron tunnels from (say) layer $s = 1$ to layer $s = 2$, the charge densities in each layer evolve dynamically following the charge injection, and at times much longer than the inverse Fermi energy multiplied by the Planck constant the differential charge density $\rho_{\sigma, s}(\r,t)$ is much smaller in magnitude than $en_\sigma$.  We focus everywhere on such long-time charge-spreading processes.  Equivalently, as we explain at the end of this section, one can say that we focus exclusively on bias voltages that are small compared to the typical interaction scale $e^2 \kf/\epsilon$.

At a semi-quantitative level, the functional behavior of the tunneling current can be anticipated using the following scaling arguments.  Consider the semiclassical process in which the injected charge $e$ of a tunneled electron spreads radially outward from the site of injection.  At some time $t$ after injection, the charge distribution $\rho(\r)$ has a typical radius $r$ and an associated energy $U(r)$.  For example, when Coulomb interactions are strong and unscreened, $U(r)$ is given by the Coulomb self-energy of the spreading charge, $U(r) \sim e^2/\epsilon r$.  The spatial gradient of energy $d U/d r$ can be said to drive the charge spreading. (In the Coulomb-dominated case this gradient is precisely the electric field).  

As the charge spreads radially outward, its energy $U(r)$ declines, and at some $r = r^*$ its energy becomes equal to the energy $eV$ associated with the bias voltage. This state with $r = r^*$ can be considered the final state of the virtual process (the ``classically-allowed state"), with an energy equal to that of the initial state before the electron tunneling.  One can estimate the charge-spreading action $S$ as the action associated with spreading of the charge packet from a small size $r_0$, which is of the order of the Fermi wavelength, to $r = r^*$.  This action depends in general on the conductivity of the CF liquid, which determines the growth rate $dr/dt$ of the charge packet.

For a CF liquid, it is important to distinguish between the \emph{physical} conductivity $\hat{\sigma}$ and the \emph{composite Fermion conductivity} $\hat{\sigma}_\tn{CF}$.  The difference between the two arises because of the electric field $\ecs$ associated with the internal CS gauge field.  (Here we follow the notation of Ref.\ \onlinecite{SimonReview}.)  Ignoring the contribution from the different spin-components for the moment, a current density $\j$ of flux-carrying CFs gives rise to a gauge electric field $\ecs(\j)$ that is perpendicular to the current.  The CFs respond to both the physical electric field $\mathbf{E}$ and the CS electric field $\ecs(\j)$, so that in the absence of density gradients one can define $\hat{\sigma}_\tn{CF}$ by
\beq
\j = \hat{\sigma}_\tn{CF}~(\mathbf{E} + \e(\j)),
\eeq 
where
\begin{align} 
\e(\j) & = \rhocs \j, \nonumber \\
\hat\rho_{CS} &= \alpha \left(\begin{array} {cc} 0 & 1 \\ -1 & 0 \end{array}\right),
\label{eq:rhocs}
\end{align}
and $\alpha = 2 \pi \hbar \phi/e^2$.  The corresponding \emph{physical} conductivity is defined by 
\beq
\j = \hat{\sigma} \E,
\eeq
so that 
\beq 
\hat{\sigma} = \left( \hat{\sigma}_\tn{CF}^{-1} + \rhocs \right)^{-1}.
\eeq
The matrix $\hat{\sigma}_\tn{CF}$ is diagonal (i.e. $[\scf]_{xy}=[\scf]_{yx}=0$). In our analysis below we use the usual assumption that $\alpha^2 [\scf]_{xx} [\scf]_{yy} \gg 1$  \cite{SimonReview}.  The validity of this inequality is discussed in detail in the Appendix.  If the electric potential is assumed to have a wave vector $q \ll \kf$ in the $x$ direction, then the physical conductivity in the $x$ direction is
\beq 
\sigma_{xx} \simeq \frac{1}{\alpha^2 [\scf]_{yy}}.
\label{eq:sigmaphys}
\eeq
This equation implies that increasing the CF conductivity $[\scf]_{yy}$ leads to a \emph{reduction} in the physical conductivity $\sigma_{xx}$, and therefore to a slower charge spreading and a larger charge-spreading action.  Heuristically, one can think that a large CF conductivity leads to a large component of current transverse to the applied electric field, and therefore to a CS field $\e$ with a component that nearly cancels the applied field $\mathbf{E}$.

For a CF system with two spin components, \cite{SimonReview}
\beq 
[\scf]_{yy} = \frac{e^2}{2 \pi \hbar} \frac{k_{F\ua} + k_{F\da}}{q} = \frac{e^2 \kf}{2 \pi \hbar q} g(\z),
\eeq
where $g(\z) = \sqrt{(1+\z)/2} + \sqrt{(1-\z)/2}$ is a monotonically decreasing function of $\z$, so that $[\scf]_{yy}$ decreases with increasing spin polarization and the physical conductivity $\sigma_{xx}$ increases.

In discussing the scaling behavior of the charge-spreading action, one can think that the typical value of $q$ is $\sim 1/r$, where $r$ is the radius of the spreading charge.  The typical value of the physical conductivity is therefore $\sigma_{xx}(r) \sim e^2/[\hbar \kf r g(\z)]$.  This conductivity defines the radial current density $j_r \sim \sigma_{xx}(r) [(1/e) dU/dr]$ associated with the spreading charge, which by continuity is related to the change in the radius by $j_r \sim (e/r^2) dr/dt$.  Thus we arrive at
\beq
\frac{dr}{dt}  \sim r^2 \frac{\sigma_{xx}(r)}{e^2} \frac{dU}{dr} \sim \frac{r}{\hbar \kf g(\z)} \frac{dU}{dr}.
\label{eq:drdtscaling}
\eeq

One can now discuss the different regimes for the charge-spreading action by considering the dominant contributions to the energy $U(r)$ that drives the charge spreading.  When the interlayer separation $d$ is large (regime I), the energy of the charge packet in a given layer is dominated by its Coulomb self-energy and is unaffected by the charge in the opposite layer.  Thus, in this limit $U(r) \sim e^2/\epsilon r$. Equation (\ref{eq:drdtscaling}) then gives a growth rate $dr/dt \propto 1/r$, so that at long times the typical size of the spreading charge is $r(t) \propto \sqrt{t}$.  The classically-allowed radius $r^*$ is given by $U(r^*) \sim eV$, so that $r^* \sim e/(\epsilon V)$, and the corresponding charge-spreading time is $t^* \propto (r^*)^2$.  The magnitude of the charge-spreading action can be estimated as the typical energy multiplied by the typical charge-spreading time,
\beq
S \sim U(r^*) t^*.
\eeq
Making this substitution for regime I gives a charge spreading action
\beq 
S_\text{I} \sim \hbar \frac{e^2 \kf/\epsilon}{e V} g(\z).
\eeq
Thus, in regime I the tunneling conductance vanishes at small voltage as $I \sim \exp[-V_0/V]$, with $V_0 \sim (e \kf/\epsilon)g(\z)$.  This result was first derived for $\z = 1$  by He, Platzman, and Halperin \cite{HPH}.  Importantly, while previous studies considered the case of full spin polarization, the simple scaling argument presented here shows that the tunneling current should \emph{increase} with increasing spin polarization, due to the rising physical conductivity $\sigma_{xx}$.

\begin{figure}
\begin{center}
\includegraphics[width=0.9\columnwidth]{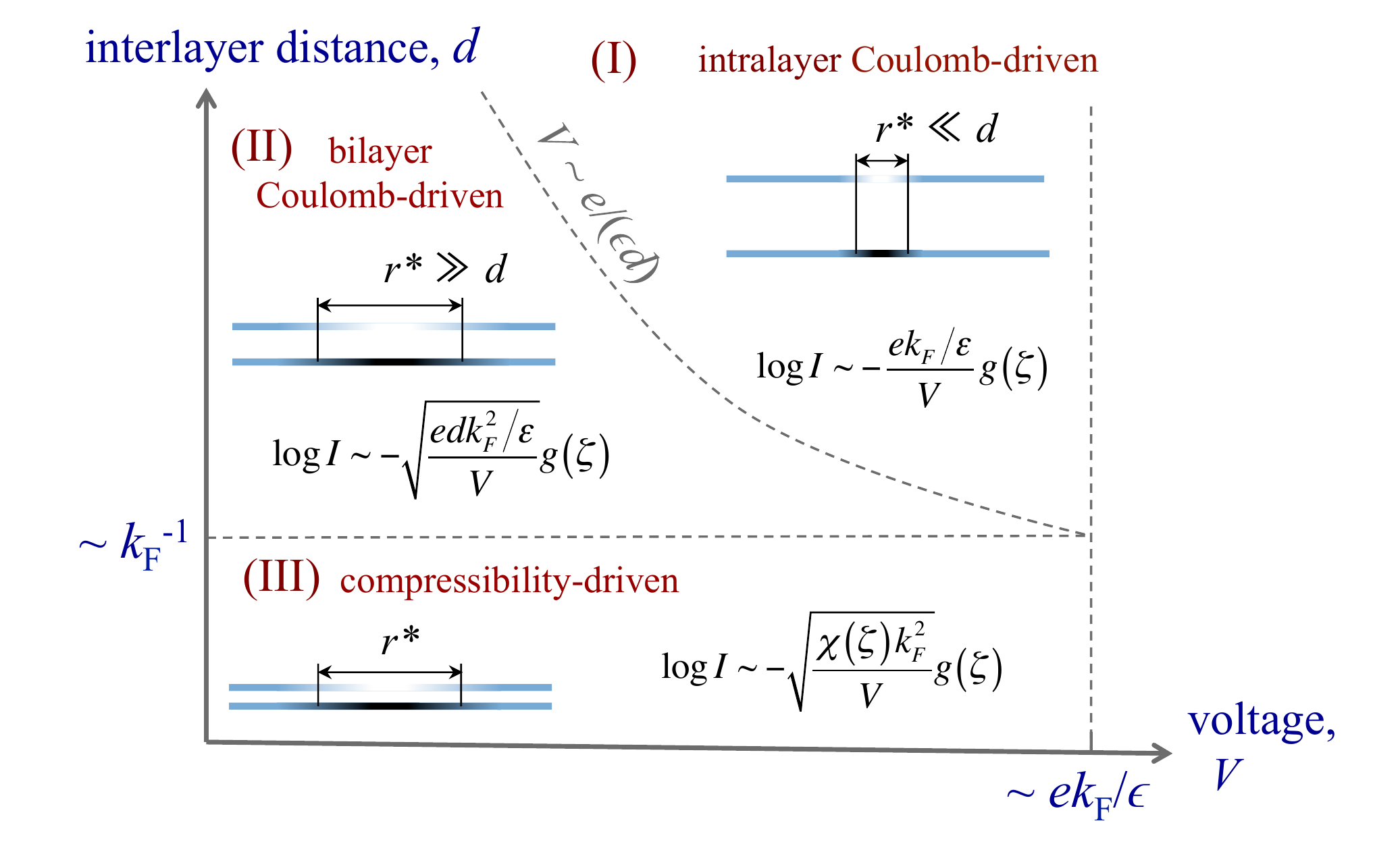}
\end{center}
\caption{Schematic map of the three regimes described in this work (labeled I, II, and III, respectively). The inset in each regime depicts the typical size $r^*$ of the spreading charge relative to the interlayer separation $d$.  In regime (I) the charge spreading is driven by the $~e^2/r^*$ Coulomb energy of the charge within each layer.  In regime (II) the Coulomb energy is reduced due to attraction of positive and negative charges in opposite layers, and the Coulomb energy of the spreading charge has the form $~e^2 d/(r^*)^2$.  In regime (III) the energy associated with the finite quantum compressibility of the spreading charge, $\sim \chi/(r^*)^2$, dominates over the Coulomb energy.}
\label{phasediag}
\end{figure}

At sufficiently small interlayer distance $d$ or sufficiently low voltage $V$, the radius $r^*$ becomes larger than the interlayer separation $d$ (regime II).  In this case the Coulomb energy of the spreading charge is affected by the attractive interaction between the opposite-sign spreading charge in the two layers, and at $r \gg d$ one can estimate $U(r)$ as the energy of a plane capacitor with charge $e$ and area $r^2$, so that $U(r) \sim e^2 d/(\epsilon r^2)$.  Equation (\ref{eq:drdtscaling}) then gives a radius that grows as $r(t) \sim t^{1/3}$ --- more slowly than in the case of an unscreened interaction.  Calculating the radius $r^*$ from $U(r^*) \sim eV$ and the corresponding time $t^*$ gives a charge-spreading action
\beq 
S_\text{II} \sim \hbar \sqrt{ \frac{e^2 d \kf^2/\epsilon }{e V} } g(\z).
\eeq
The corresponding functional form $I \sim \exp[-\text{const.}/\sqrt{V}]$ of the tunneling current was first pointed out in Refs.\ \cite{HPH, XGW94}.  Below we provide a detailed exploration of this regime, including the effect of finite spin polarization.  However, one can see already from the arguments above that the tunneling current in both regimes I and II is expected to increase with increasing spin polarization.

Finally, one can consider the regime where $d$ is so small that the mean-field Coulomb energy of the spreading charge is effectively eliminated due to the close proximity of the two layers (as mentioned above, we still assume that there is no instability to excitonic condensation).  In this regime the energy of the spreading charge is dominated by the residual, short-ranged component of the interactions that give the CF liquid its finite compressibility.  In other words, the CF liquid has a finite thermodynamic density of states $\chi^{-1} = dn/d\mu$, where $\mu$ is the chemical potential, and this finite thermodynamic density of states produces an outward diffusive current. The energy associated with the compressibility is $U(r) \sim \chi / r^2$.  For the CF problem, where the compressibility arises from the short-ranged component of interactions (which are not quenched even when the interlayer spacing is small), $\chi$ is of order $e^2 /(\epsilon \kf)$.  The dependence of $\chi$ on the spin polarization $\z$ cannot be predicted \textit{a priori}.  As we show below, its value depends on Landau parameters.

It is important to note that the compressibility $\chi$ does not depend on the spin of the initially injected electron [as depicted in Fig.\ \ref{cartoon}(b)].  As the packet of charge density spreads outward, the resulting CS electric field creates currents of both spin components, thereby mixing the spin densities spatially.  After a long time the spin composition of the spreading charge is determined only by the compressibilities of the two spin components, and it does not necessarily reflect the spin of the injected electron or the spin polarization of the unperturbed ground state.  A careful derivation of $\chi$ in terms of the compressibilities $dn_{\ua}/d \mu_{\ua}$, $dn_{\da}/d \mu_{\ua}$, and $dn_{\da}/d \mu_{\da}$ is presented below.

Accepting $\chi(\zeta)$ as a phenomenological parameter, one can use Eq.\ (\ref{eq:drdtscaling}) as before to make an estimate of the charge-spreading action.  This procedure gives
\beq 
S_\text{III} \sim \hbar \sqrt{ \frac{\chi(\z) \kf^2}{e V} } g(\z).
\eeq
Thus, the dependence of the conductivity on spin polarization is altered by the spin-dependence of the compressibility. 

In order to understand the crossover between the three regimes, one can simply equate $U(r^*)$ for regimes I and II, and regimes II and III.  This gives the two dashed lines $V \sim e/(\epsilon d)$ and $d \sim \kf^{-1}$ drawn in Fig.\ \ref{phasediag}. The maximum voltage for applicability of our semiclassical description can be estimated by demanding that the typical size $r^*$ of the spreading charge is much longer than the Fermi wavelength $\kf^{-1}$.  Since the voltage is related to $r^*$ by $eV \sim U(r^*)$, our description applies only when $eV \ll U(\kf^{-1})$.  For regime I, this inequality amounts to $V \ll e \kf/\epsilon$.  For regime II, the inequality is equivalent to $V \ll e \kf^2 d/\epsilon$, which is automatically satisfied within the boundaries of regime II.  For regime III, our description applies only when $V \ll \chi \kf^2/e$.  Since $\chi$ is of order $e^2/(\epsilon \kf)$, the condition for applicability in regime III is the same as in regime I, $V \ll e \kf/\epsilon$.  This boundary is marked as the vertical dashed line in Fig.\ \ref{phasediag}.

\begin{widetext}

\section{Hydrodynamic description of charge spreading}
\label{sec:hydro}

In the hydrodynamic description, the evolution of the charge currents $\j_{\sigma, s}(\r)$ and the charge densities $\rho_{\sigma,s}(\r)$ are described using semiclassical equations of motion. The charge-spreading action is the action associated with the evolution of the charge densities and charge currents in the wake of the tunneling event.  Following Ref.\ \onlinecite{LS}, our approach to calculating the charge-spreading action is to write down the equations of motion, and then to write the simplest quadratic action that reproduces the known equations of motion.  Once this action is known, we can solve for $\j$ and $\rho$ associated with a ``bounce path", in which a charge $e$ is inserted at time $-\tau$ and removed at a later time $\tau$.

\subsection{Hydrodynamic equations}

In the hydrodynamic description, one assumes that there is a local equilibrium with a well defined electrochemical potential $\Phi(\r)$, which varies smoothly with position on the scale of the Fermi wavelength and slowly with time relative to the inverse Fermi energy. In our case the electrochemical potential has both a spin and layer index, and can be written
\beq
\Phi_{\sigma,s}(\r) = e\vp_{s}(\r) + \mu_{\sigma,s}(\r),
\label{Phi}
\eeq
where $\mu_{\sigma,s}(\r)$ denotes the chemical potential associated with the compressibility and $\vp_{s}(\r)$ is the electric potential at position $\r$ in layer $s$.  In the remainder of this subsection we suppress the layer index `$s$' in all equations; the electric potential, the densities, and the currents are understood to correspond to the same layer.  The CF current is given by
\begin{subequations}
\beq
\j_{\sigma} &=& \hat\sigma_{\tn{CF},\sigma} \left(-\frac{1}{e}\nabla \Phi_{\sigma} + \ecs(\j) \right),
\label{eq:jdelPhia}\\
\j &=& \sum_\sigma ~\j_\sigma,
\label{eq:jdelPhib}
\eeq
\end{subequations}
where $\hat\sigma_{\tn{CF},\sigma}$ and $\e$ are the CF conductivity matrix (for spin component $\sigma$) and the CS electric field, respectively.
Let us write $\nabla \Phi$ as
\beq
\nabla \Phi_{\sigma} =- e\E + \frac{1}{e}\sum_{\sigma'}~\chi_{\sigma\sigma'} \nabla \rho_{\sigma'},
\eeq
where $\E = -\nabla\varphi$ is the physical electric field and we have defined
\beq 
\chi_{\sigma\sigma'} \equiv \frac{d \mu_\sigma}{d n_{\sigma'}},
\eeq
where $\rho_\sigma = e n_\sigma$. The coefficients $\chi_{\sigma\sigma'}$ define the spin-selective compressibilities of the system, and are related to Landau parameters in a way that we explain below. For the moment we leave them as unspecified parameters.

Let us now assume that the electric field is in the $x-$direction, which is also assumed to be the direction along which the density perturbation has a larger density gradient (i.e. $|\nabla_x \rho| \gg |\nabla_y \rho|$~). Under these assumptions, the $x-$component of the current in Eq.\ (\ref{eq:jdelPhia}) is given by
\beq
\frac{j_\ua^{x}}{[\sigma_{\tn{CF},\uparrow}]_{xx}} - \alpha (j_\ua^y + j_\da^y) &=& 
 E_x - \frac{1}{e^2} \bigg(\chi_{\ua\ua}\nabla_x\rho_\ua + \chi_{\ua\da} \nabla_x\rho_\da\bigg)  \nonumber \\ 
\frac{j_\da^{x}}{ [\sigma_{\tn{CF},\downarrow}]_{xx}} - \alpha (j_\ua^y + j_\da^y) &=& 
 E_x - \frac{1}{e^2} \bigg(\chi_{\da\da}\nabla_x\rho_\da + \chi_{\da\ua} \nabla_x\rho_\ua\bigg).
\label{eq:junsimplified}
\eeq

Examining the $y-$component of the current in Eq.\ (\ref{eq:jdelPhia}) gives $j_\ua^y + j_\da^y = - \alpha \sigma_{yy}^T (j_\ua^x + j_\da^x)$,
where we have defined
\beq
\sigma^T_{yy} &=& [\sigma_{\tn{CF},\ua}]_{yy} + [\sigma_{\tn{CF},\da}]_{yy}.
\eeq
We can substitute for $j_\ua^y + j_\da^y$ in Eq.\ (\ref{eq:junsimplified}) and apply the inequality $\alpha^2 ~ \sigma^T_{yy} \gg 1/[\sigma_{\tn{CF},\sigma}]_{xx}$ to each of the spin components to arrive at the following set of equations:
\beq
\alpha^2 \sigma_{yy}^T (j_\ua^x + j_\da^x) &=& 
 E_x - \frac{1}{e^2} \bigg(\chi_{\ua\ua}\nabla_x\rho_\ua + \chi_{\ua\da} \nabla_x\rho_\da\bigg)  \nonumber \\ 
\alpha^2 \sigma_{yy}^T (j_\ua^x + j_\da^x) &=& 
E_x - \frac{1}{e^2} \bigg(\chi_{\da\da}\nabla_x\rho_\da + \chi_{\da\ua} \nabla_x\rho_\ua\bigg).
\label{eq:jsimplified}
\eeq
The justification for the inequalities $\alpha^2 ~ \sigma^T_{yy} \gg 1/[\sigma_{\tn{CF},\sigma}]_{xx}$ is discussed in detail in the Appendix.  We note here only that the inequality is fully justified in regimes I and II, while in regime III it is marginal at worst.

From Eq.\ (\ref{eq:jsimplified}) one can immediately see that the gradient terms in the two equations are equal. Thus, the ratio of the up and down spin densities of the evolving charge perturbation is given by
\beq 
\bigg(\chi_{\ua\ua} - \chi_{\da\ua} \bigg)~ \rho_{\ua} = \bigg(\chi_{\da\da} - \chi_{\ua\da}\bigg)~\rho_{\da}
\label{eq:rhoupdownratio}
\eeq
(note that the distributions $\rho_\ua(\r)$ and $\rho_\da(\r)$ must be normalized and therefore cannot differ by an additive constant).
Physically, this relation arises because the CS field mixes the two spin components until they are in local equilibrium with each other, which guarantees that their density gradients satisfy Eq.\ (\ref{eq:rhoupdownratio}). Thus, even if (say) an up-spin CF is injected into the system, the CS field quickly induces the evolving density perturbation to develop a mixture of both up and down-spin CF components that reflects their thermodynamic compressibilities. Equivalently, one can say that the electrochemical potential $\Phi_\sigma$ in Eq.\ (\ref{Phi}) that governs the current flow becomes independent of the spin $\sigma$. It is then possible to write down a single hydrodynamic equation, independent of the spin components.

Using the constraint in Eq.\ (\ref{eq:rhoupdownratio}), the hydrodynamic equation (\ref{eq:jsimplified}) can be simplified to
\beq
j_x = \sigma_{xx}~\bigg[E_x - \frac{\chi_{\tn{eff}}}{e^2} \nabla_x~\rho\bigg],
\label{totalj}
\eeq
where $j_x = j_\ua^x + j_\da^x$ and $\sigma_{xx} = 1/(\alpha^2\sigma^T_{yy})$ is the physical conductivity, as before.  We have introduced an effective coefficient
\beq
\chi_{\tn{eff}} = \frac{\chi_{\ua\ua}\chi_{\da\da} - \chi_{\ua\da} \chi_{\da\ua}}{(\chi_{\ua\ua} - \chi_{\da\ua}) + (\chi_{\da\da} - \chi_{\ua\da})}.
\label{effchi}
\eeq
From now on, we shall assume $\chi_{\ua\da} = \chi_{\da\ua}$.

The derivation of Eq.\ (\ref{totalj}) illustrates that, at long wavelengths and in the long time limit, one can write the hydrodynamic equations in terms of the \emph{total} current and \emph{total} density, summed over both spin components.  In this way the spinful problem is reduced to a spinless problem, written in terms of the physical conductivity and a renormalized compressibility. The effective compressibility $\chi_{\tn{eff}}$ can be written in terms of Landau parameters, as we demonstrate below. Further, we show in Sec.\ \ref{sec:instanton} (and as in Ref. \cite{DCBSPL1}), that $\chi_\text{eff}$ is the same as the thermodynamic $d\mu/dn = (2 \pi \phi)^2 \chi_d$.  It is worth emphasizing that while it may seem obvious that the effective compressibility for the total charge is given by $\chi_{\tn{eff}} = d\mu/dn$, its appearance in the hydrodynamic equations is in fact a nontrivial result that comes from the influence of the CS field.  The usual thermodynamic expression for $d\mu/dn$ is obtained by assuming equilibration between the two spin components, so that a given perturbation $\delta n$ of density can be divided between the two spin sectors in a way that minimizes the total energy.  In our problem, however, there are no processes which can flip the CF spin.  Instead, mixing of spin densities happens through the influence of the CS field, even when the chemical potential of one spin species is independent of the density of the other and $\chi_{\ua \da} = 0$.  Over long length scales, this mixing produces a ratio of spin densities given by Eq.\ (\ref{eq:rhoupdownratio}), which leads to an effective compressibility that is equal to the thermodynamic one. 

The electric field $\E$ in our problem should be calculated self-consistently from the evolving charge-distribution,
\beq
\E(\r,t) = -\nabla_\r \int_{\r'}~\rho(\r',t)~V(|\r-\r'|),~\tn{where}\nonumber \\
V(|\r-\r'|) = \left[ \frac{1}{\epsilon |\r - \r'|} - \frac{1}{\epsilon \sqrt{(\r - \r')^2 + d^2}} \right],
\eeq
and $\rho(\r) = \sum_\sigma \rho_\sigma(\r)$ is the total charge density. Here we have exploited the symmetry between the two layers, which have equal and opposite charge densities $\rho(\r)$ at any given time. 

We now obtain the various spin-selective compressibilities, defined in the limit where the mean-field Coulomb energy is effectively quenched but there is still a residual interaction on short length scales between the different spin components of the CFs. We describe the partially spin-polarized composite Fermi liquid phenomenologically within a Landau Fermi liquid approach by introducing Landau parameters \cite{PN}. 
Assuming rotational invariance, we introduce the dimensionless Landau parameters $F_\ell^{\sigma\sigma'} = \sqrt{m_\sigma^* m_{\sigma'}^*} f_\ell^{\sigma\sigma'}/(2\pi)$. We restrict ourselves to only the $\ell = 0$ component, corresponding to the compression mode of the Fermi surfaces. Following Landau's expansion to quadratic order, the energy is given by \cite{SimonReview}
\beq
\delta {\cal{E}}(\rho_\ua, \rho_\da) = \pi \frac{(1+F_0^{\ua\ua})}{m_\ua^*} \frac{\rho_\ua^2}{e^2} + \pi \frac{(1+F_0^{\da\da})}{m_\da^*} \frac{\rho_\da^2}{e^2} + 2\pi\frac{F_0^{\ua\da}}{\sqrt{m_\ua^* m_\da^*}} \frac{\rho_\ua \rho_\da}{e^2}.
\eeq
(From here onward we set $\hbar = 1$.)

The individual compressibilities are then given by,
\beq
\chi_{\ua\ua} &=& \frac{2\pi(1+F_0^{\ua\ua})}{m_\ua^*},\\
\chi_{\da\da} &=& \frac{2\pi(1+F_0^{\da\da})}{m_\da^*},\\
\chi_{\ua\da} &=& \frac{2\pi F_0^{\ua\da}}{\sqrt{m_\ua^* m_\da^*}},
\eeq
where $F_0^{\ua\da} = F_0^{\da\ua}$. Plugging in the explicit form of these quantities, the effective coefficient in Eq.\ (\ref{effchi}) is given by
\beq
\chi_{\tn{eff}} = \frac{\pi}{2}\bigg[\frac{2}{m_{\tn{eff}}} + \frac{F_0^{s\ua} + F_0^{s\da}}{2\pi} - \frac{\bigg(\frac{1}{m^*_{\ua}} - \frac{1}{m^*_{\da}} + \frac{F_0^{s\ua} - F_0^{s\da}}{2\pi}\bigg)^2}{\bigg(\frac{2}{m_{\tn{eff}}} + \frac{F_0^{a}}{2\pi}\bigg)}\bigg],
\label{chieff}
\eeq
where $F_0^{s\ua(\da)} = F_0^{\ua\ua(\da\da)}+F_0^{\ua\da}$ and $F_0^a = F_0^{\ua\ua} + F_0^{\da\da} - 2F_0^{\ua\da}$. Here we have introduced a {\it reduced} mass, $m_{\tn{eff}} = 2m^*_\uparrow m^*_\downarrow / (m^*_\uparrow + m^*_\downarrow)$ (the factor of $2$ ensures that in the limit of identical masses, $m_{\tn{eff}}=m^*_{\uparrow(\downarrow)})$. As mentioned above, $\chi_{\tn{eff}} = d\mu/dn = (2\pi\phi)^2~\chi_d$ (see Sec.\ \ref{sec:instanton} and Eq. ({\color{blue} 20}) of Ref.\ \cite{DCBSPL1}). Moreover, in the familiar limit of $F_0^{\ua\ua} = F_0^{\da\da}$ and $m_\ua^*=m_\da^*$, we recover $\chi_{\tn{eff}}\sim(1+F_0^s)/m^*$ \cite{PN}.

In addition to the hydrodynamic equation, Eq.\ (\ref{totalj}), the current and density are related by the continuity equation
\beq
\frac{d \rho}{dt} + \nabla \cdot \j = {\cal{J}}(\r,t),
\label{eq:continuity}
\eeq
where ${\cal{J}}(\r,t)$ represents a source associated with injection of an external charge.  In our problem,
\beq
{\cal{J}}(\r,t) = e \delta(\r) \left[ \delta(t+\tau) - \delta(t-\tau)\right](1 - 2\delta_{s,2}),
\eeq
which describes the insertion of a positive (negative) charge on layer $1$ ($2$) at time $t = - \tau$, and its subsequent removal at time $t = + \tau$.

Using Eqs.\ (\ref{totalj}) and (\ref{eq:continuity}), one can solve for both the density and the current as a function of frequency and momentum,
\beq
\rho(\omega,q) = \frac{{\cal{J}}(\omega)}{|\omega| + q^2~\sigma_{xx}(q) [V(q) + \chi_{\tn{eff}}/e^2]},
\label{eq:rhoTsol}
\eeq
and
\beq
j(\omega,q) = -iq~\sigma_{xx}(q)\bigg[V(q) + \frac{\chi_{\tn{eff}}}{e^2} \bigg]~\rho(\omega,q),
\label{eq:jTsol}
\eeq
where ${\cal{J}}(\omega) = 2ie\sin(\omega\tau)~(1 - 2\delta_{s,2})$.

\subsection{The form of the charge-spreading action}

Following Ref.\ \onlinecite{LS}, the action associated with this system can be determined by writing down the simplest quadratic action in $\rho$ and $\j$ that correctly reproduces the hydrodynamic equations.  In Fourier space, this action (per layer) is
\beq 
S_{\tn{hydro}} = \frac{1}{2} \sum_{i\omega_n}\int_\q  \bigg[K^{-1}(\q,\omega) |j(\omega,\q)|^2 + M(\q,\omega) |\rho(\omega,\q)|^2 \bigg],
\label{eq:hydroaction}
\eeq
where $K^{-1}(\q,\omega)$ and $M(\omega,\q)$ are as yet undetermined functions and $i\omega_n$ are Bosonic Matsubara frequencies. The above action is to be supplemented with the continuity equation, Eq.\ (\ref{eq:continuity}). 

We demand that the above action reproduce the equations of motion (i.e. the hydrodynamic equation) in Eq.\ (\ref{totalj}), which fixes
\beq
K(\q,\omega) &=& \sigma_{xx}(q)~|\omega|,\\
M(\q,\omega) &=& V(q) + \frac{\chi_{\tn{eff}}}{e^2}. 
\label{KM}
\eeq
An equivalent way of arriving at the same conclusion is as follows: for the action defined in Eq.\ (\ref{eq:hydroaction}), the current-current correlation function $\langle \j(\omega,\q)~\j(-\omega,-\q)\rangle = K(\q,\omega)$, which is by definition given by $\sigma_{xx}(q)~|\omega|$. On the other hand, the coefficient of $|\rho(\omega,\q)|^2$ in the action contains the contribution from the Coulomb energy and the finite compressibility, which are represented by the two terms in $M(\q,\omega)$.

Inserting the solutions for $\rho(\omega,\q$) and $j(\omega,\q)$ derived above [Eqs.\ (\ref{eq:rhoTsol}) and (\ref{eq:jTsol})] for the specific boundary condition ${\cal{J}}(\omega)$ into Eq.\ (\ref{eq:hydroaction}), one arrives at the following expression for the action in each layer:
\beq
S_{\tn{hydro}}(\tau) = \frac{1}{2}\sum_{i\omega_n}\int_\q \frac{|{\cal{J}}(\omega)|^2}{|\omega|~} \frac{V(q) + \chi_{\tn{eff}}/e^2}{|\omega| + q^2~\sigma_{xx}(q) [V(q) + \chi_{\tn{eff}}/e^2]}.
\label{eq:Shydroint}
\eeq
[See Eq.\ (\ref{actMCS}) for the analogous action obtained using the instanton-based approach \cite{XGW94,DCBSPL1}.]

The total action, after subtracting the action associated with the work done by the voltage source, is given by
\beq
S_\text{tot} =  S_{\tn{hydro}}(\tau) - 2 e V \tau.
\label{eq:Stot}
\eeq
The charge-spreading time $\tau_*$ associated with a particular voltage is found by minimizing the action with respect to $\tau$. The tunneling conductivity is then of the form $\sim\tn{exp}[-S_\tn{tot}(\tau_*(V))]$.

Below we evaluate this expression for different limiting cases of the Coulomb interaction potential $V$, corresponding to regimes I, II, and III outlined above.

\subsubsection{Regime I: Coulomb-driven charge spreading at large layer separation}
\label{hydI}
In the limit of large layer separation and relatively small voltage, such that $q d \gg 1$, the Coulomb interaction $V(q) \simeq 2 \pi/(\epsilon q)$ dominates over the term $\chi_{\tn{eff}}/e^2$ associated with the compressibility in the limit of $q\rightarrow0$. Thus, one can ignore the corrections due to a finite $\chi_{\tn{eff}}$ in the action of Eq.\ (\ref{eq:Shydroint}).

As before, the tunneling conductivity is proportional to $\tn{exp}[-{\cal{S}}(\tau_*(V))]$, where $\tau_*(V)$ is the characteristic charge spreading time. In regime $\rm{I}$, this is
\beq
S(\tau_*(V)) = 2~A~g(\z) \frac{e^2/\epsilon l_B}{2eV},
\label{eq:SI}
\eeq
where $A=4\pi$ and the extra numerical prefactor of $2$ is for the contribution from the two layers and we have set $\phi=2$.

\subsubsection{Regime II: Coulomb-driven charge spreading at small layer separation}
\label{hydII}
In the limit where the layer separation is small enough and the typical size of the spreading charge is large enough that $q d \ll 1$, the Coulomb interaction saturates to a constant value $V(q) \simeq 2 \pi d / \epsilon$.  In this limit one can take the term $M(\q,\omega) = V(q) + \chi_{\tn{eff}}/e^2$ in Eq.\ (\ref{eq:Shydroint}) to be a constant independent of $\q$ and $\omega$; we simply denote it as $M$. One then arrives at a total action
\beq
S(\tau_*(V)) =2~C~ \kf g(\z) \sqrt{\frac{Me^2}{2eV}},
\label{eq:StotsmallV}
\eeq
where $C=(-2^6\Gamma^3(-1/3)/3^7\pi)^{1/2}$ and the extra numerical prefactor of $2$ corresponds to the contribution from the two layers.

If the interlayer spacing remains large enough that $e^2 d/\epsilon \gg \chi_{\tn{eff}}/e^2$, then Eq.\ (\ref{eq:StotsmallV}) reduces to
\beq
S(\tau_*(V)) = 2~C ~\kf g(\z) \sqrt{ \frac{2\pi e^2 d/\epsilon}{2 e V} } .
\label{eq:SII}
\eeq

\subsubsection{Regime III: Compressibility-driven charge spreading}

Finally, if $d$ is so small that $\chi_{\tn{eff}}/e^2 \gg e^2 d/\epsilon$, then the charge-spreading action is dominated by the compressibility.  In this limit the final result for the charge-spreading action becomes
\beq
S(\tau_*(V)) = 2~C ~ \kf g(\z) \sqrt{ \frac{\chi_{\tn{eff}}}{2 e V} } .
\label{eq:SIII}
\eeq

In this regime the dependence of the tunneling current on the spin polarization $\zeta$ depends on the way in which the Landau-parameters in $\chi_{\tn{eff}}$ [see Eq.\ (\ref{chieff})] vary with $\zeta$. This dependence is of course not known {\it a priori}. As we describe in Sec.\ \ref{sec:discussion}, however, this dependence can be deduced from experiments. In principle, it is possible that our description of the charge spreading action in this regime can correctly explain the experimental results of Ref. \cite{JPE16} when this $\z$-dependence is taken into account. 

\section{Electronic spectral function}
\label{sec:instanton}
In this section we briefly review the computation of the electron Green's function in the partially spin-polarized quantum Hall bilayers at total filling $\nu=1/2$ in each layer, which we presented in an accompanying paper \cite{DCBSPL1}.  We compare the corresponding results with those obtained from our semiclassical analysis in the previous two sections. As mentioned in Sec.\ \ref{sec:qualitative}, generalizing from the result for a spin-polarized system \cite{NB93} we attach flux to electrons (of either spin orientation) such that a CF of any given spin orientation sees $\phi$ flux quanta attached to electrons of {\it both} spin components only in the same layer. This amounts to the transformation
\beq
\s_{s,\sigma}(\r) =  \s_{e,s,\sigma}(\r)~\tn{exp}\bigg[i\phi\int_{\r'} \tn{arg}(\r-\r')~ n_{s}(\r') \bigg],
\eeq
where $\s_{e,s,\sigma}(\r)$ and $\s_{s,\sigma}(\r)$ represent the electron and CF annihilation operators, respectively, at position $\r$ in layer $s$ with spin quantum number $\sigma$. As before, $n_s(\r)$ is the total density of electrons (or, equivalently, CFs within the HLR theory) in layer $s$. For $\phi=2$, the filling $\nu = \nu_\uparrow + \nu_\downarrow=1/2$ in each layer and the CFs of either spin orientation do not see any magnetic field on average. The resulting Fermi surfaces of the CF have Fermi wave vectors $k_{F\ua(\da)}$, as denoted in Eq.\ (\ref{kf}) \cite{JKJspin}.  

The low-energy field theory for the CF Fermi surfaces minimally coupled to the gauge field is then given by \cite{HLR,NB93}
\beq
\label{CFlag}
\L &=& \L_0 + \L_{\tn{int}} + \L_{\tn{CS}},\\
\L_0 &=& \sum_{s,\sigma} \bigg(\s_{s,\sigma}^\dagger(\r,\tau)[\d_\tau + i a_0^s(\r,\tau)]\s_{s,\sigma}(\r,\tau) + \frac{1}{2m^*_\sigma}\s_{s,\sigma}^\dagger(\r,\tau) [-i\nabla+\D \a^s(\r,\tau)]^2 \psi_{s,\sigma}(\r,t)  \bigg),\nonumber\\
\L_{\tn{int}} &=& \sum_{s,s'} \frac{1}{2}\int_\r \int_{\r'} e^2 V_{s,s'}(\r-\r') :n_s(\r) n_{s'}(\r'): 
\nonumber
\eeq
where $m^*_{\uparrow(\downarrow)}$ denote the effective masses for the different spin-components, $\D\a$ denotes the gauge field minus $e \A$, with $\A$ being the external vector potential, and `$:~:$' denotes normal ordering. The Coulomb interaction, $V_{s,s'}(\r) = 2\pi/\left(\epsilon \sqrt{r^2 + d^2(1-\delta_{s,s'})}\right)$, is insensitive to the spin label. The Chern-Simons term is given by
\beq
\L_{\tn{CS}} &=& -\frac{i}{2\pi}\sum_{ss'} \int_{\r} K^{-1}_{ss'} ~a_0^s(\r,\tau)~ \hat{z}\cdot[\nabla\times \a^{s'}(\r,\tau)],
\label{CS}
\eeq
where $K_{ss'}$ is diagonal with respect to the layer index: $K_{ss'} = \phi~ \delta_{ss'}$. 

We are interested in computing the single-{\it electron} Green's function that corresponds to tunneling an electron with spin $\sigma$ into layer $s$ at $\r=0$ and time $t=0$ and then removing an electron at $\r=0$ with the same spin and from the same layer at a later time $t=2\tau$,{\footnote{The interval is chosen to be $2\tau$ such that it agrees with the setup in the hydrodynamic description in Section\ \ref{sec:hydro}.}} which is given by $G_{s,\sigma}(\tau) =\langle \psi_{e,s,\sigma}(\vec{0},2\tau)~\psi_{e,s,\sigma}^\dagger(\vec{0},0)\rangle$,
\beq
G_{s,\sigma}(\tau) =  \int {\cal{D}}[\psi~a]~ \psi_{s,\sigma}(2\tau)~\psi_{s,\sigma}^\dagger(0)~\delta(\M) ~\tn{exp}(-S[\psi^\dagger,\psi,a_\mu]).
\label{Gst}
\eeq
Here, $S[\psi^\dagger,\psi,a_\mu]$ is the imaginary-time action corresponding to the field theory introduced in Eq.\ (\ref{CFlag}). For the fully spin-polarized case, this calculation was done in Ref.\ {\cite{XGW94}}. It is clear that the electron Green's function is different from the CF Green's function, and $\delta(\M)$ denotes precisely the boundary condition in space-time on the gauge field, which involves creating and annihilating two flux quanta. The path integral measure ${\cal{D}}[\psi~a] \equiv \prod_{s',\sigma'}~D\psi_{s',\sigma'}^\dagger~D\psi_{s',\sigma'}~Da^{s'}_\mu$. 

Formally, we can integrate out the CFs and obtain an effective action purely in terms of the gauge-fields, $S_{\tn{eff}}[a_\mu]$. It is then reasonable to assume that the low-energy suppression of the spectral function is dominated by the exponential saddle point contribution of this Maxwell-Chern-Simons action, with momentum- and frequency-dependent dielectric/permeability functions (inherited from the gapless CF Fermi surfaces), in the presence of the above boundary condition. 

For the bilayer problem, the boundary condition translates to the creation of a monopole in the top and an anti-monopole in the bottom layer at time $t=0$, both of which are removed at a later time $2\tau$ at the same position $\r=0$. In the limit of times much longer than the inverse Fermi energy, this process couples only to the low-energy diffusive mode \cite{HLR,NB93} with $\omega\sim V(q) q^3$, where $V(q) = 2\pi(1-e^{-qd})/(\epsilon q)$. However, as we discussed in Ref.\ \onlinecite{DCBSPL1}, the inserted monopole/antimonopole does not have a spin quantum number and the magnetization associated with the spreading charge may quickly evolve to contain a mixture of both components that may not reflect the magnetization $\z$ of the background. When the charge spreading is driven purely by the Coulomb energy of the perturbation, the magnetization of the perturbation is irrelevant for the charge spreading, since the Coulomb interaction is independent of spin. However, this is not the case in the regime where the dominant energy scale driving the charge spreading is provided by the finite compressibility of the CF fluid. In this case, as was discussed also in the previous section, in the long-time limit the magnetization of the perturbation is determined by the ratio of the different spin compressibilities.

Within a random phase approximation (RPA) treatment of the effective action \cite{HLR} in Eq. (\ref{CFlag}), we obtain $S_\tn{eff}[a] = S_\tn{em} + S_{\tn{CS}}$, where
\beq
S_\tn{em} = \frac{1}{2}\sum_{i\omega_n}\int_\q \bigg[ \ve(\q,\omega) |\e_{\q,\omega}|^2 + \beta(\q,\omega) |b_{\q,\omega}|^2\bigg],
\label{effact}
\eeq 
where 
$e_\alpha = \d_0 a_\alpha - \d_\alpha a_0$ is the electric field and $b^s = (\d_x a_y^s - \d_y a_x^s)$ is the magnetic field associated with the internal gauge field in layer $s$. The coefficients are given by
\beq
\ve(\q,\omega) &=& \frac{2(k_{F\uparrow} + k_{F\downarrow})}{4\pi |\omega_n| q} = \frac{2k_F}{4\pi |\omega_n| q}~g(\zeta),\\
\beta(\q,\omega) &=& \chi_d + \frac{e^2}{(2\pi\phi)^2} V(\q),\eeq
where $g(\zeta) = \sqrt{(1+\zeta)/2} + \sqrt{(1-\zeta)/2}$ and, as described in Eq.\ (\ref{chieff}), $\chi_d = (\partial\mu/\partial n)/(2\pi\phi)^2 =\chi_{\tn{eff}}/(2\pi\phi)^2$, where $\mu$ is the chemical potential.

It is worth emphasizing that the action written in Eq.\ (\ref{eq:hydroaction}) is equivalent to the ``electromagnetic'' action, $S_{\tn{em}}$, in Eq.\ (\ref{effact}). The $|\j|^2$ term in Eq.\ (\ref{eq:hydroaction}) indicates the action associated with the current, which is equivalently represented in terms of the gauge electric field $|\e|^2$, such that the conductivity is associated with the effective dielectric function $\varepsilon(\vec{q}, \omega)$.  Similarly, the $|\rho|^2$ term in Eq.\ (\ref{eq:hydroaction}) is associated with the density, which is equivalent to the $|b|^2$ term in the electromagnetic action above, with the effective inverse magnetic permeability $\beta(\vec{q}, \omega)$ playing the role of the $M(\q,\omega)$ term in Eq.\ (\ref{KM}).

Following Refs.\ \cite{Tru93,XGW94,DCBSPL1}, and for the boundary conditions described above, the action is given by
\beq
S_{\tn{eff}}(\tau) = \int_\omega \int_\q \frac{(1-\cos(2\omega\tau))}{|\omega|}\frac{V(q) + (2\pi\phi)^2 \chi_d/e^2}{|\omega| + q^2\bigg[\frac{q}{2\pi\phi^2\kf g(\zeta)}\bigg]\bigg(V(q) + (2\pi\phi)^2 \chi_d/e^2 \bigg)}.
\label{actMCS}
\eeq
Remarkably, this action is identical to the hydrodynamic action, $S_{\tn{hydro}}(\tau)$, obtained in Eq. \ref{eq:Shydroint}, even though it is obtained using a completely different approach. All of the regimes that we discussed in our hydrodynamic analysis can therefore be recovered in a straightforward fashion from the above action.  

\end{widetext}

\section{Summary and Outlook}
\label{sec:discussion}

In this paper we have presented a derivation of the action associated with tunneling of electrons between quantum Hall bilayers in situations where each layer is at a compressible filling. In an accompanying paper \cite{DCBSPL1}, we explicitly computed the electron Green's function using an instanton-based approach. In the present paper, we have focused primarily on describing the same action within a hydrodynamic formulation, where the specific form of the conductivity at long wavelengths serves as an input. 
Our main interest was understanding the functional dependence of the tunneling current on voltage, interlayer separation, and spin polarization.  Our results are summarized in Fig.\ \ref{phasediag}, and in Eqs.\ (\ref{eq:SI}), (\ref{eq:SII}), and (\ref{eq:SIII}).  We find that both the hydrodynamic and instanton-based approaches give identical results.

In light of the recent experiment \cite{JPE16}, one of our most striking results is that previous descriptions where the mean-field Coulomb energy is responsible for the charge spreading (as in Refs.\ \cite{XGW94, LS, HPH}) cannot account for the experimental observations of Ref.\ \onlinecite{JPE16}.  In particular, such descriptions yield a tunneling current that \emph{increases} with increasing spin polarization, while the experiment observed the opposite trend.  This discrepancy has led us to identify a new regime of behavior, denoted regime $\rm{III}$, in which the two layers are sufficiently closely spaced that the mean-field Coulomb energy is quenched, and the charge spreading is driven instead by the finite compressibility of the CF liquid. 
Within this regime, the dependence of the TA on the spin polarization is indirectly governed by the dependence of the Landau parameters (which determine the compressibility) on the polarization [see Eq.\ (\ref{chieff})]. Future experiments can check this dependence explicitly by measuring the inverse compressibility as a function of spin polarization using capacitance or field penetration measurements \cite{Eis94}, which allow one to extract $\chi_d\sim \chi_{\tn{eff}}$. Moreover, if our proposed explanation for the observation of Ref.\ \cite{JPE16} is correct, the tunneling current should have the functional form $\ln I \propto - 1/\sqrt{V}$ at small $d/\lb$. In the future it may also be interesting to study how the TA evolves into the zero-bias conductance peak associated with the onset of exciton condensation \cite{EisReview} as one crosses below the critical value of $d/\lb$.  
 
On the other hand, as the parameter $d/\lb$ is increased, the system moves away from the compressibility-dominated regime and into the Coulomb-dominated regimes. If the behavior of $\chi_\text{eff}(\z)$ within the compressibility-dominated regime is indeed consistent with the experimentally observed decrease in tunneling current with increasing spin polarization, then the magnitude of this decrease must weaken as the value of $d/\lb$ is increased.  Indeed, within the Coulomb-dominated regimes at $d/\lb \gg 1$ the dependence of the TA on the spin-polarization must go in the opposite direction as observed in the experiments of Ref.\ \cite{JPE16}.  One can therefore use our theory to predict that as a function of increasing $d/\lb$ the dependence of tunneling current on $\z$ should reverse sign. While studying a wide range of $d/\lb$ can be challenging experimentally, we note that one can also access the compressibility-dominated regime by placing a metallic layer in close proximity to a single compressible quantum Hall layer.  In this situation the metallic layer can effectively screen out the long-range Coulomb interactions, leaving only the compressibility of the quantum Hall system to drive the charge spreading.

Finally, it is worth noting that while the HLR theory of the spin-polarized $\nu=1/2$ state has been remarkably well supported by many experiments \cite{Willett1,Kang,Goldman}, recent years have seen a surge of interest in alternate theoretical descriptions of the $\nu = 1/2$ state.  In particular, a well known concern with the HLR formulation of the theory is the absence of particle-hole symmetry, which should exist in the lowest Landau level in the limit of large magnetic field. A recent proposal attempts to resolve this concern by postulating that the CFs are {\it Dirac} fermions \cite{DTS}, such that the physical action of particle-hole transformation acts as time-reversal symmetry on the Dirac fermions. While a microscopic derivation of this proposal is currently lacking, a number of works have contributed to the ongoing efforts to resolve this puzzle \cite{Mross,Max,Senthil}. However, it has also been recently pointed out in Ref.\ \cite{Chong} that when response functions are properly evaluated within HLR theory, there is an emergent particle-hole symmetry.  

So far, the many experiments that were seen to be in agreement with the predictions of the original formulation of the HLR theory are also consistent with the revised formulation in terms of the Dirac-CF theory. Therefore, new experiments are necessary to clearly distinguish between the two scenarios. Unfortunately, the low bias TA that we are considering here is unlikely to be able to distinguish between the two scenarios. Since the TA is determined only by the low-energy properties near the CF Fermi surface, which are identical within the two scenarios, we expect that results for the charge-spreading action are also identical within the two formulations.  

\acknowledgements
We thank J. P. Eisenstein and I. Sodemann for useful discussions. We acknowledge the hospitality of IQIM-Caltech, where a part of this work was completed. DC is supported by a postdoctoral fellowship from the Gordon and Betty Moore Foundation, under the EPiQS
initiative, Grant GBMF-4303, at MIT. BS was supported as part of the MIT Center for Excitonics, an Energy Frontier Research Center funded by the U.S. Department of Energy, Office of Science, Basic Energy Sciences under Award no. DE-SC0001088. PAL acknowledges support by DOE under Award no. FG02-03ER46076. DC acknowledges the hospitality of the Aspen Center for Physics, which is supported by NSF grant PHY-1607611.


\appendix
\section{Validity of the inequality $\alpha^2 [\scf]_{xx} [\scf]_{yy} \gg 1$}

For a compressible CF system, the conductivity $\hat{\sigma}_\tn{CF}(q, \omega)$ at finite frequency $\omega$ can be calculated within RPA \cite{HLR, SimonReview}.  At low enough frequency that $\omega/(q v_\tn{F}) \ll 1$, and in the absence of impurity scattering, this conductivity to leading order in $\omega/(q v_F)$ is given by \cite{SimonReview}
\begin{align}
    [\scf]_{xx} & \simeq \frac{-i e^2 \kf \omega}{2 \pi \hbar v_\tn{F} q^2}, \\
    [\scf]_{yy} & \simeq \frac{e^2 \kf}{2 \pi \hbar q}.
\end{align} 
Here, $v_\tn{F} = \hbar \kf/m^*$ denotes the Fermi velocity, and the wave vector $q$ is taken to be in the $x$ direction, as above. In the typical experimental situation, one can usually assume the strong inequality $\alpha^2 [\scf]_{xx} [\scf]_{yy} \gg 1$, where $\alpha = 2 \pi \hbar \phi/e^2$, for the typical magnitudes of the corresponding conductivities. This inequality implies that the CFs respond primarily to the CS electric field, rather than the physical electric field. In our problem, where the typical frequency $\omega$ and wave vector $q$ are determined by the process of charge spreading, the use of this inequality should be justified carefully.  For simplicity, in this appendix we discuss the spinless case ($\zeta=1$), while in the main text we adapt the inequality to the case of partial spin polarization. 

We first note that, since the effective mass of the CFs arises from their short-ranged interactions, the typical mass scale is such that $e^2 \kf/\epsilon \sim \hbar^2 \kf^2/m^*$.  This implies that $m^* \sim \epsilon \hbar^2 \kf/e^2$, or $v_\tn{F} \sim e^2/(\epsilon \hbar)$.  Making this substitution into $[\scf]_{xx}$ implies that the inequality $\alpha^2 [\scf]_{xx} [\scf]_{yy}\gg1$ is equivalent to
\beq
\frac{\epsilon \hbar \phi^2 \kf^2 \omega}{e^2 q^3} \gg 1. \nonumber
\eeq
In our problem, the typical wave vector associated with the charge spreading is $1/r^*$, and the typical frequency is $1/t^*$.  So the necessary inequality is
\beq
\frac{\epsilon \hbar \phi^2 \kf^2 (r^*)^3}{e^2 t^*} \gg 1.
\label{eq:inequality}
\eeq
We now consider whether this inequality is satisfied in each of the three regimes summarized in Fig.\ \ref{phasediag}.

In regime I, where the energy scale that drives the charge spreading is $U(r) \sim e^2/(\epsilon r)$, the size of the spreading charge follows $r^2 \sim e^2 t/(\epsilon \hbar \kf)$ [see Eq.\ (\ref{eq:drdtscaling})].  So Eq.\ (\ref{eq:inequality}) becomes $\phi^2 \kf r^* \gg 1$.  Since we are considering charge spreading processes over length scales much longer than the Fermi wavelength $\kf^{-1}$, the inequality is satisfied.

In regime II, the mean field Coulomb energy is $U(r) \sim e^2 d/(\epsilon r^2)$, and consequently the size of the spreading charge evolves according to $r^3 \sim d e^2 t/(\epsilon \hbar\kf)$.  Thus, Eq.\ (\ref{eq:inequality}) becomes $\phi^2 \kf d \gg 1$.  Since the condition $d \gg \kf^{-1}$ is already part of the definition of regime II, the inequality is again satisfied.

Finally, in regime III, the charge spreading is driven by the energy scale $\chi/r^2$ associated with the finite compressibility, and $r^3 \sim \chi  t/(\hbar \kf)$.  The inequality of Eq.\ (\ref{eq:inequality}) therefore becomes $\phi^2 \epsilon \chi \kf/e^2 \gg 1$. The magnitude of the compressibility is of order $\chi\sim \hbar^2/m^* \sim e^2/(\epsilon \kf)$, multiplied by a combination of Landau parameters (as discussed in Sec.\ \ref{sec:hydro}) that may be numerically large. Even in the worst-case scenario where we set all the Landau parameters to zero, the inequality we are considering becomes equivalent to $\phi^2 \gg 1$. When the number $\phi$ of attached fluxes is not too large, such as in the half-filled Landau level where $\phi = 2$, this inequality is marginal. 
Nonetheless, even in the worst-case scenario where the inequality is only marginally satisfied, our primary results for regime III are unaltered.  In particular, the charge spreading is still driven by the finite compressibility at $d \ll \kf^{-1}$, with a functional dependence $\log I \propto -1/\sqrt{V}$ and with a spin polarization-dependence that is affected by Landau parameters.

\bibliographystyle{apsrev4-1_custom}
\bibliography{qhe}
\end{document}